\RequirePackage{fix-cm}
\documentclass[twocolumn]{svjour3}          
\smartqed  
\usepackage{graphicx}
\usepackage{epstopdf} 
\usepackage{epsfig}
\usepackage{amsmath}
\usepackage{bbding}
\usepackage{color}
\usepackage{amsfonts}

\begin{document}

   \title{Systematic search for islets of stability in the standard map for large parameter values}

	\author{Alexandre R. Nieto \and Rub\'{e}n Cape\'{a}ns \and Miguel A.F. Sanju\'{a}n 
	}

	\institute{A. R. Nieto \and R. Cape\'{a}ns \and M. A. F. Sanju\'{a}n \at Nonlinear Dynamics, Chaos and Complex Systems Group, Departamento de
		F\'{i}sica, Universidad Rey Juan Carlos, Tulip\'{a}n s/n, 28933 M\'{o}stoles, Madrid, Spain \\
		\email{alexandre.rodriguez@urjc.es} \\
		\email{ruben.capeans@urjc.es} \\
		\email{miguel.sanjuan@urjc.es} \\	
	}
	
	\date{Received: date / Accepted: date}
	
	\maketitle

	
	\begin{abstract}

	In the seminal paper (Phys.~Rep.~\textbf{52}, 263, 1979), Boris Chirikov showed that the standard map does not exhibit a boundary to chaos, but rather that there are small islands (``islets") of stability for arbitrarily large values of the nonlinear perturbation. In this context, he established that the area of the islets in the phase space and the range of parameter values where they exist should decay following power laws with exponents $-2$ and $-1$, respectively. In this paper, we carry out a systematic numerical search for islets of stability and we show that the power laws predicted by Chirikov hold. Furthermore, we use high-resolution 3D islets to reveal that the islets' volume decays following a similar power law with exponent $-3$.
		

	\end{abstract}
	
	\section{Introduction} \label{sec:Introduction}
	
	The standard map, also known as Chirikov-Taylor standard map, is a paradigmatic 2D area-preserving map given by \cite{Chirikov}
	\begin{equation} 
		\begin{aligned}
			\theta_{n+1} &= \theta_n + p_{n+1},\\
			p_{n+1} &= p_n + K\sin\theta_n,
		\end{aligned} 
	\end{equation}
	where $p_{n}$ and $\theta_n$ are taken modulo $2\pi$; and $K>0$ is a constant whose role is to increase the nonlinear perturbation. The physical meaning of $\theta$ and $p$ depends on the system under consideration.
	
	This is a quite general model which describes the dynamics of a nonlinear oscillator under periodic perturbations, and it has been broadly studied in the context of nonlinear dynamics. Some real-world physical situations described by the standard map include a kicked rotor \cite{Zaslavsky} and the motion of a charged particle in a magnetic bottle \cite{Chirikov}.  
	
	For small values of $K$ (say $K<4$), the phase space $(\theta,p)$ is occupied by a chaotic sea and a main Kolmogorov-Arnold-Moser (KAM) island that covers a significant area of the phase space. It is noteworthy that all the KAM islands in this system surround stable periodic orbits. The coexistence of stable and chaotic dynamics is a hallmark of Hamiltonian systems and, in words of Zaslavsky, ``is one of the most striking and wonderful discoveries ever made"~\cite{Zaslavsky}. As $K$ increases, the area of the KAM island is irregularly reduced until being eventually destroyed for $K\approx7$. For higher values of the parameter $K$, a fully chaotic regime might be expected. However, small regions (``islets") of stability emerge in an approximate periodic manner, even for arbitrarily large parameter values \cite{Duarte94}. Akin to KAM islands, islets of stability are regions filled with KAM tori where orbits are dynamically trapped. However, islets are not directly related to the main KAM island. As a matter of fact, they do not emerge in bifurcations of the main family of periodic orbits. Instead, they appear in saddle-node bifurcations out in the chaotic sea \cite{Barrio09}. Although islets have not received much attention in the literature, their existence is not a unique feature of the standard map, but they also appear in general Hamiltonian systems and area-preserving maps \cite{Barrio20,Nieto23,Mackay}.  
	
	Chirikov himself suggested that the length of the parameter interval $\Delta K$ where the islets exist and their area $A$ in the phase space should decay following power laws like $\Delta K \propto K^{a_1}$ and $A\propto K^{a_2}$. The predicted exponents are $a_1=-1$ and $a_2=-2$. The latter exponent was numerically investigated by Contopoulos~\textit{et al.} in $2005$~\cite{Contopoulos}. They classified the islets in two different types and obtained a decay exponent ($-2.45$ and $-2.3$) for each type. Although these numbers are relatively close to the theoretical value $-2$ (relative error about $15-20\%$), they do not provide definitive numerical evidence supporting the idea that these laws govern the decay of magnitudes in the system. To the best of our knowledge, there have not been more refined estimations of these exponents.
	
	In this paper, we conduct rigorous numerical simulations to validate the accuracy of Chirikov's predictions, which we find to be remarkably precise. In particular, we obtain the predicted decay exponents with relative errors of less than $0.6\%$. Moreover, we extend the previous predictions to the case of the volume in the $(\theta,p,K)$ space, obtaining an exponent $-3$.
	
	The manuscript is organized as follows. First, in Section~\ref{sec2}, we introduce a systematic approach for computing islets of stability. In Section~\ref{sec3}, we explore numerous islets across a wide range of parameters, demonstrating that their length, area, and volume strictly follow power laws. Finally, in Section~\ref{sec4}, we summarize the key findings and discuss the potential implications of the results.
	
	\section{Islets of stability}\label{sec2}
	
	A general overview of the standard map dynamics can be obtained simply by plotting orbits in the phase space. However, since we aim to locate islets of stability in a systematic manner, we need a method to detect the last KAM curve, i.e., the boundary between a KAM island and the chaotic sea. In this section, we present the methods we have implemented to accurately locate the islets. 
	
	Due to the modulo operation, the standard map is a mapping from the two-dimensional torus to itself. Therefore, any orbit is confined to the phase space region $[0,2\pi]\times[0,2\pi]$. Nevertheless, orbits within the chaotic sea will eventually visit the entire phase space, except the region bounded by the last KAM curve. Using this simple fact, and following Sanjuán \textit{et al.}~\cite{Sanjuan}, we define regions (``leaks") in the chaotic sea. If an orbit never falls into them, we classify it as a part of a KAM island. Naturally, for this method to work the leaks must be located in an area that is known \textit{a priori} to be in the chaotic sea. In particular, we define a left leak $L_1\equiv[0.1\pi,0.3\pi]\times[0,2\pi]$ and a right leak $L_2\equiv[1.7\pi,1.9\pi]\times[0,2\pi]$. This choice introduces two rectangular leaks of width $0.2\pi$ that are symmetrical about $\theta=\pi$. We highlight that this choice is arbitrary and any region in the chaotic sea can be used to serve this purpose.
	
	Once the leaks are defined, we divide the phase space into a grid of initial conditions and we iterate the system for each of them. If after one or more iterations an orbit falls into one of the leaks, we stop the iteration process and we classify the corresponding initial condition as a part of the chaotic sea. On the contrary, if after a fixed maximum number of iterations the orbit has never fallen into the leaks, we classify the corresponding initial condition as a part of a KAM island. To visualize the result, we assign blue (red) color to the initial conditions whose orbits fall into the left (right) leak. Finally, we assign white color to initial conditions belonging to a KAM island.
	
\begin{figure}[h!]
		\centering
		\includegraphics[clip,height=5cm,trim=0cm 0cm 0cm 0cm]{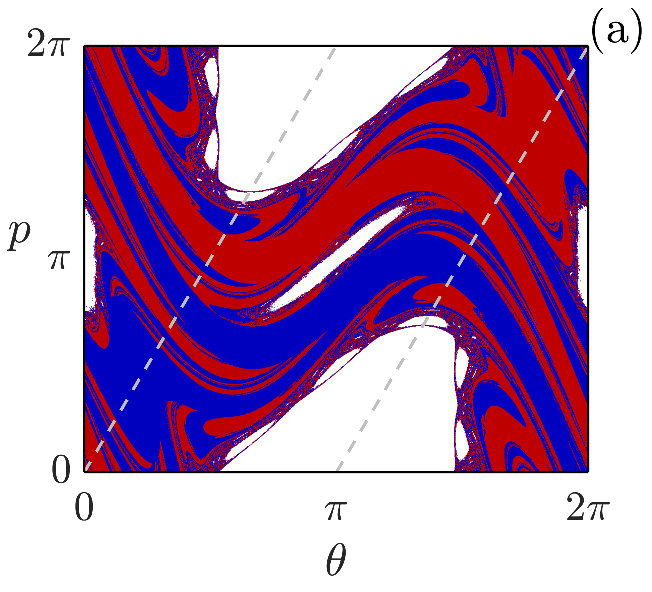}	
		\hspace{0.2cm}
		\includegraphics[clip,height=5cm,trim=0cm 0cm 0cm 0cm]{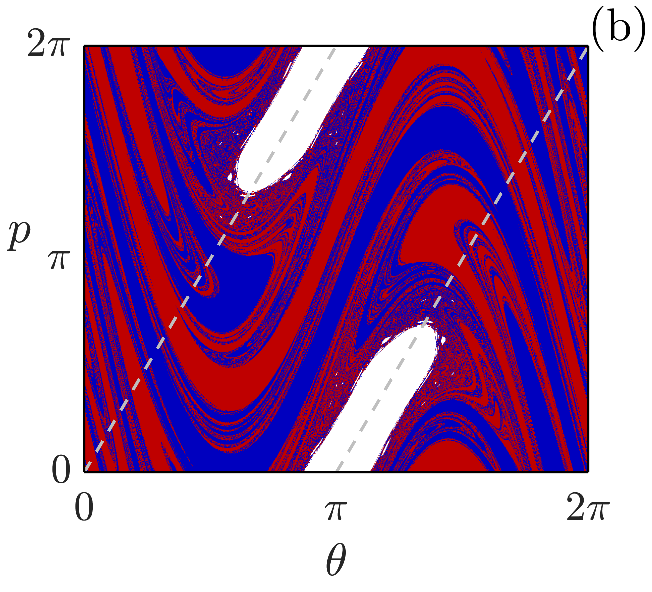}	
		\includegraphics[clip,height=5cm,trim=0cm 0cm 0cm 0cm]{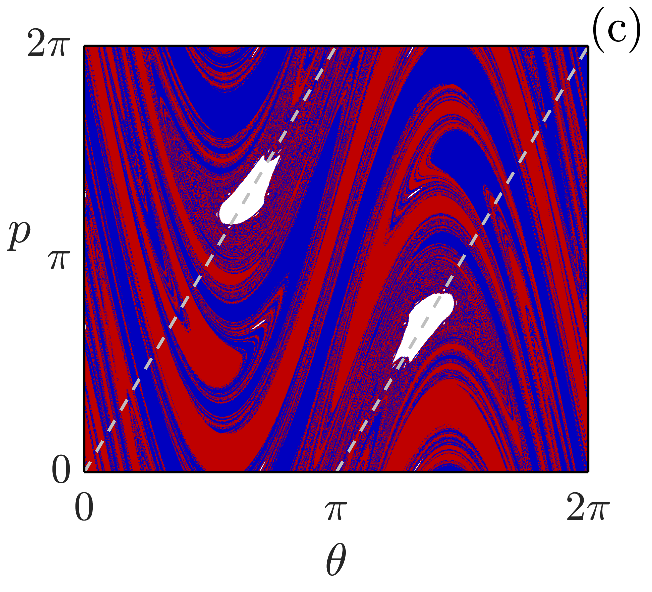}
		\hspace{0.2cm}	
		\includegraphics[clip,height=5cm,trim=0cm 0cm 0cm 0cm]{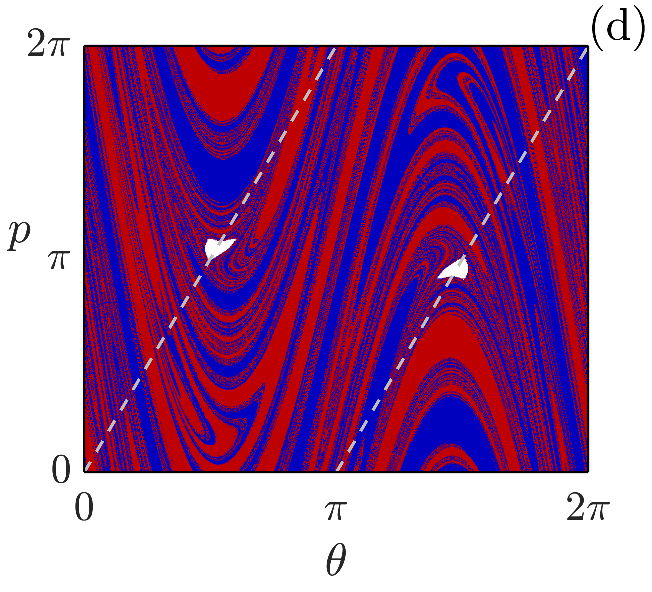}	
		\caption{Exit basin diagram in the phase space showing KAM islands (white color) for parameter values (a) $K=2$, (b) $K=4$, (c) $K=5$, and (d) $K=6$. Blue (red) color represents chaotic orbits that fall into the left (right) leak. The gray dashed lines correspond to the symmetry lines $p=2\theta$ and $p=2\theta-2\pi$.  }
		\label{Fig1}
	\end{figure}

	The procedure that we have just introduced is equivalent to the numerical simulation of exit basins in open Hamiltonian systems \cite{Contopoulos02}. For this reason, we will refer to the corresponding data and plots as ``exit basin diagrams". Some examples computed for different values of $K$ are depicted in Fig.~\ref{Fig1}. For $K<4$, the main KAM island surrounds a periodic orbit located at $(\theta,p)=(\pi,0)$ [or, equivalently, $(\theta,p)=(\pi,2\pi)$], as shown in panels (a) and (b). For higher values of $K$ [see panels (c) and (d)], the periodic orbit moves along two symmetry lines $p=2\theta$ and $p=2\theta-2\pi$, which are represented with gray dashed lines in Fig.~\ref{Fig1}. Some periodic orbits generating islets of stability also appear over these symmetry lines.
	
\begin{figure}[h!]
		\centering
		\includegraphics[clip,height=6.5cm,trim=0cm 0cm 0cm 0cm]{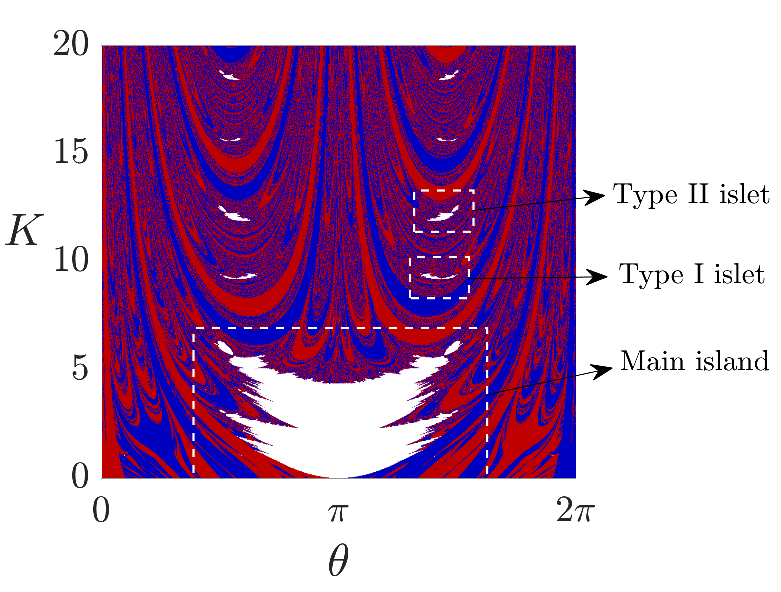}		
		\caption{Exit basin diagram showing in white color the main KAM island and some islets in the $(\theta,K)$ plane. Blue (red) color represents chaotic orbits that fall into the left (right) leak. To obtain these data, once an initial condition $\theta$ is chosen, the remaining coordinate is given by $p=2\theta-2\pi$ if $\theta>\pi$ or by $p=2\theta$ if $\theta\leq\pi$. The islets highlighted with white dashed rectangles are shown with high resolution in Fig.~\ref{Fig3} [panels (a) and (c)]. }
		\label{Fig2}
	\end{figure}

	Since islets appear in a reduced range of parameter values, it is difficult to identify them by computing exit basin diagrams in the phase space. Furthermore, as we shall show later, the interval $\Delta K$ where an islet exists is reduced as $K$ increases. Thus, this task becomes even more arduous for large parameter values. However, the fact that the main stable periodic orbits, and hence the KAM islands, lie on symmetry lines can help us to narrow our search. In particular, we can preliminarily locate islets by constructing exit basin diagrams in which we choose initial conditions along a given symmetry line for different values of $K$. As an example, in Fig.~\ref{Fig2} we show the exit basin diagram computed along the symmetry lines $p=2\theta-2\pi$ and $p=2\theta$ for parameter values in the range $K\in[0,20]$. We emphasize that other symmetry lines, such as $\theta=p$, could be used to find different islets. For low values of $K$, we observe the main KAM island, which is destroyed for $K\approx7$. This island possesses a characteristic fractal tree-like structure that is ubiquitous in area-preserving maps and Hamiltonian systems (see, for example, Refs.~\cite{Nieto23,Greene}).
	
\begin{figure*}[h!]
		\centering
		\includegraphics[clip,height=5cm,trim=0cm 0cm 0cm 0cm]{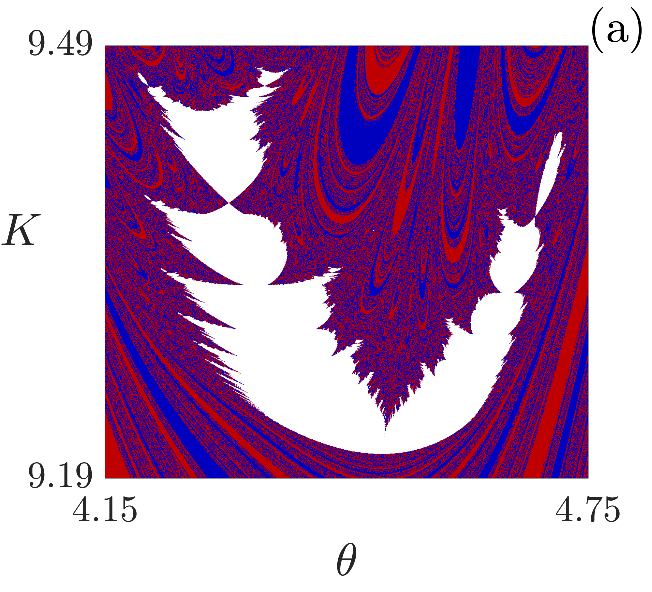}	
		\hspace{0.2cm}
		\includegraphics[clip,height=5cm,trim=0cm 0cm 0cm 0cm]{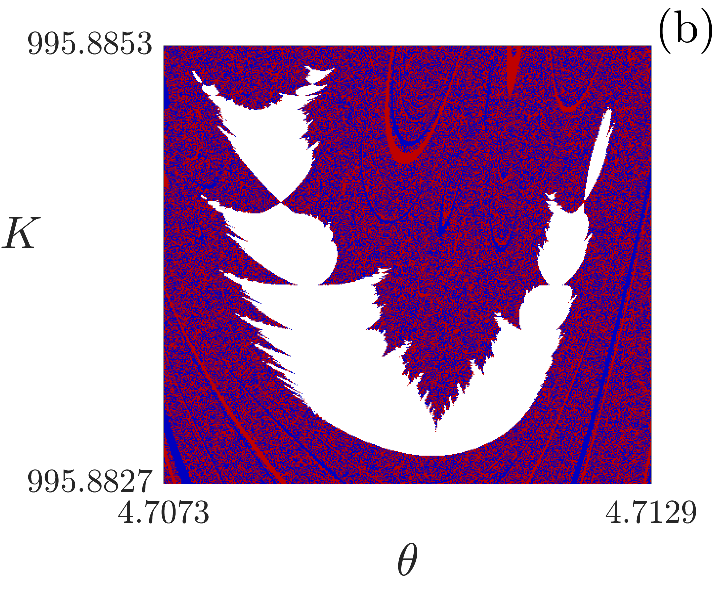}	
		\includegraphics[clip,height=5cm,trim=0cm 0cm 0cm 0cm]{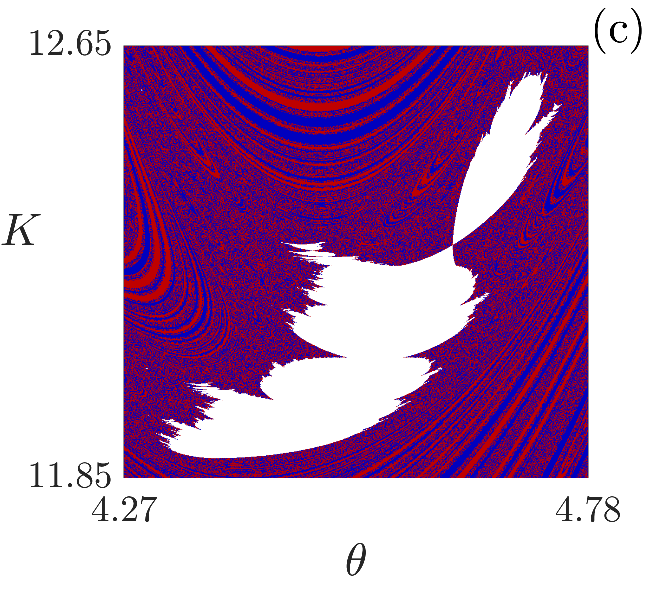}
		\hspace{0.2cm}	
		\includegraphics[clip,height=5cm,trim=0cm 0cm 0cm 0cm]{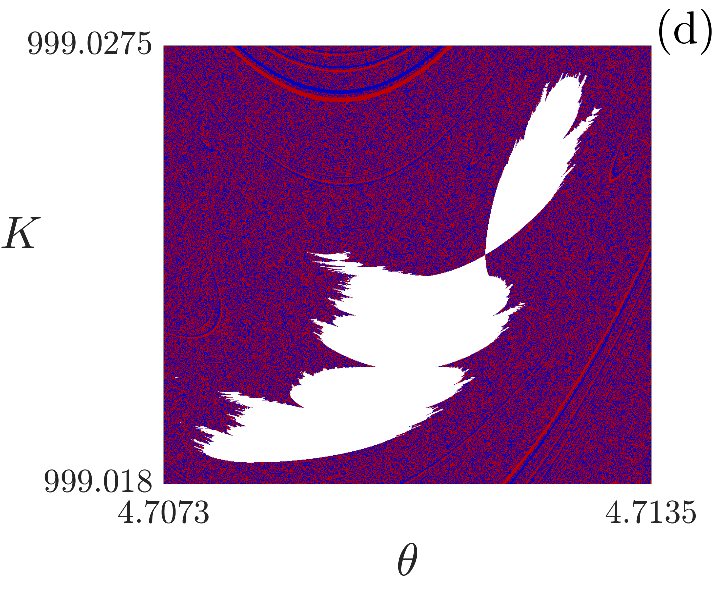}	
		\caption{Exit basin diagram in the $(\theta,K)$ plane showing islets of stability (white color) of (a,b) type I and (c,d) type II. It is clear that islets of the same type are identical scaled copies. Blue (red) color represents chaotic orbits that fall into the left (right) leak. To obtain these data, once an initial condition $\theta$ is chosen, the remaining coordinate is given by $p=2\theta-2\pi$. }
		\label{Fig3}
	\end{figure*}

	\newpage
	After the destruction of the main island, the fully chaotic regime is reached, but only momentarily. The first islet of stability appears for $K\approx 9.19$, and it is followed by a second islet appearing for $K\approx 11.85$. As it has been shown by Contopoulos \textit{et al.} \cite{Contopoulos}, to precisely study the characteristics of the islets it is necessary to classify them in different types. Each islet type is characterized by the period of its central periodic orbit and can be found in a particular symmetry line. In this work, we focus our attention on islets surrounding period-$4$ orbits (type I) and period-$2$ orbits (type II). These periodic orbits, which are periodic on the torus but nonperiodic on the cylinder, have been referred to as ``accelerator modes" in the literature \cite{Kedar99}, and their diffusion properties have received some attention \cite{Wang23,Manos14,Miguel13}. The first islet of each type, which are also highlighted with white dashed rectangles in Fig.~\ref{Fig2}, are shown with high resolution in panels (a) and (c) of Fig.~\ref{Fig3}. As can be noticed by looking at the scale of the figures, type II islets are rather bigger than type I.
	
	As $K$ is increased, a scaled copy of each type of islet appears in approximate parameter intervals of $2\pi$.  Their size is reduced as $K$ increases, but their shape remains unaltered. In other words, exact copies of the islets appear recurrently, simply changing their size as the leaves in a fern. This can be intuitively seen in Fig.~\ref{Fig2}, where it is clear that the second islets of each type are smaller. To illustrate the astonishing similarity between islets of the same type, we depict islets of types I and II for values of $K$ close to $1000$ in Fig.~\ref{Fig3}b and Fig.~\ref{Fig3}d. We can see with naked eye that their main features, ignoring their size, correspond to a great extent to the islets appearing for low values of $K$ (panels (a) and (c) of Fig.~\ref{Fig3}).

	\section{Decay of magnitudes} \label{sec3}
\begin{figure}[h!]
		\centering
		\includegraphics[clip,height=8.7cm,trim=0cm 0cm 0cm 0cm]{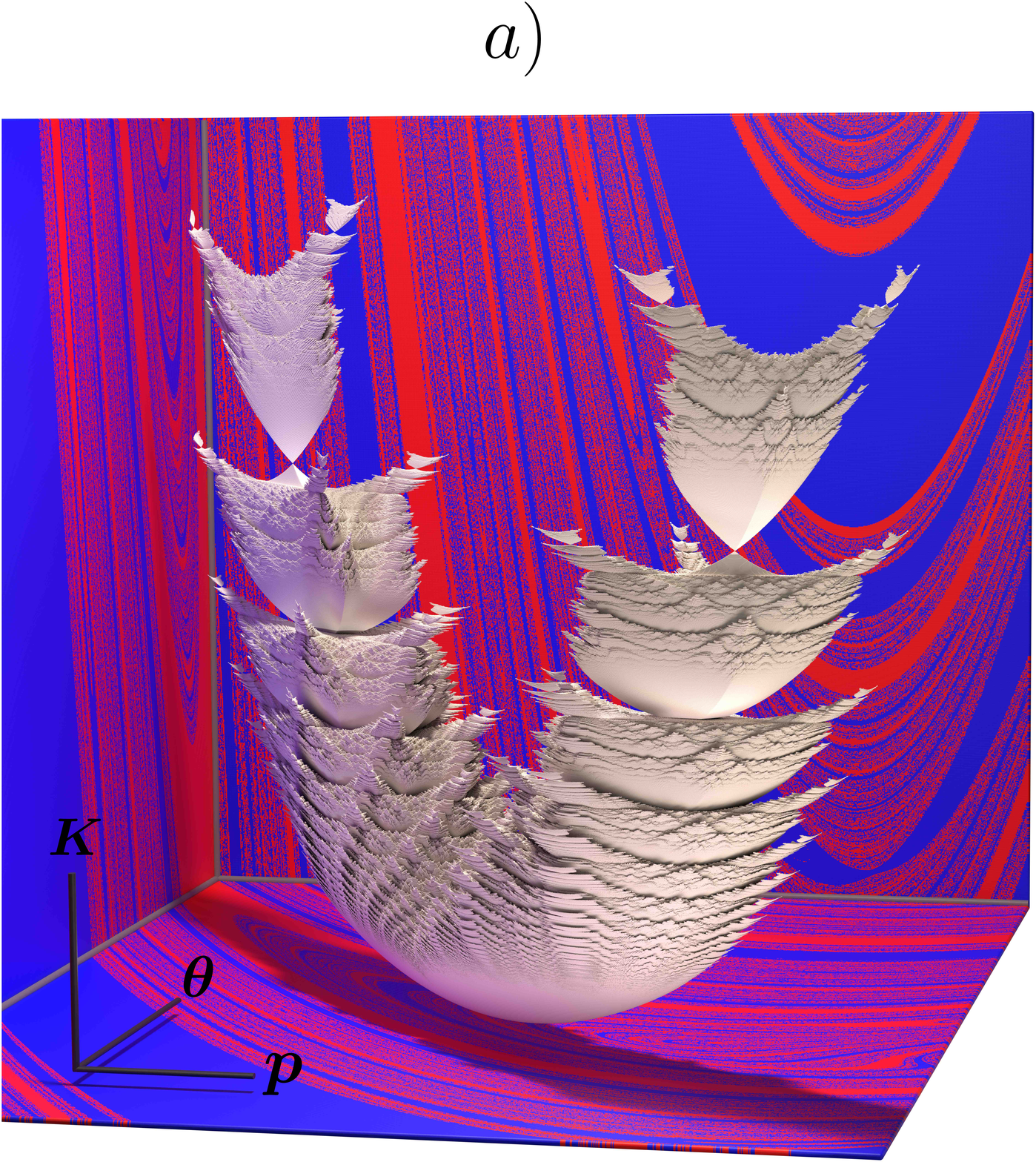}	
		\includegraphics[clip,height=8.7cm,trim=0cm 0cm 0cm 0cm]{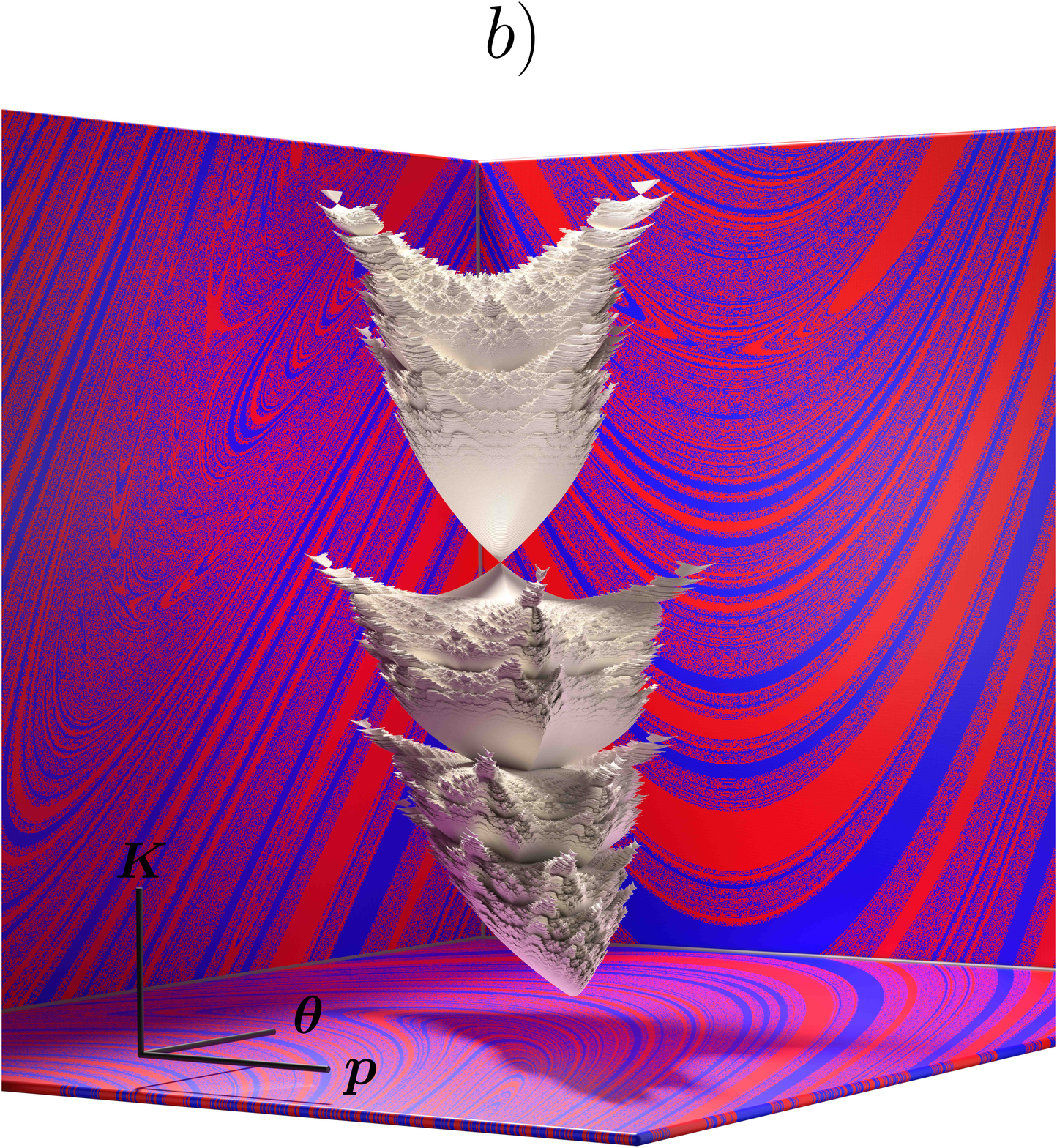}
		\caption{Islets of stability of (a) type I and (b) type II in the $(\theta,p,K)$ space. White color represents initial conditions within an islet, while blue (red) color is used on three of the cube's faces to represent chaotic orbits that fall into the left (right) leak. To properly visualize the islets, the remaining chaotic initial conditions are left transparent. Great detail of the islets arises when zooming in. }
		\label{Fig5}
	\end{figure}

	In this section, we compute the length, area and volume of the islets, as well as we obtain the exponents governing the decay of these magnitudes. To accomplish this, we systematically locate multiple islets following the procedure described in Sec. II. Afterward, we construct high-resolution 3D exit basin diagrams in the $(\theta,p,K)$ space. As an illustrative example, Fig.~\ref{Fig5} displays the first islet of each type. 
	
	These 3D representations offer a comprehensive view of the complex evolution of the islets as the parameter $K$ is increased. The rich fractal structures of the islets can be seen with great detail by zooming in the figure. From these data, a Poincaré surface of section for a fixed value of $K$ brings exit basin diagrams equivalent to those showed in Fig.~\ref{Fig1}. In a similar fashion, defining the Poincaré section $p=2\theta-2\pi$, the exit basin diagrams of Fig.~\ref{Fig3} are obtained. 
	
\begin{figure}[h!]
		\centering
		\includegraphics[clip,height=8cm,trim=0cm 0cm 0cm 0cm]{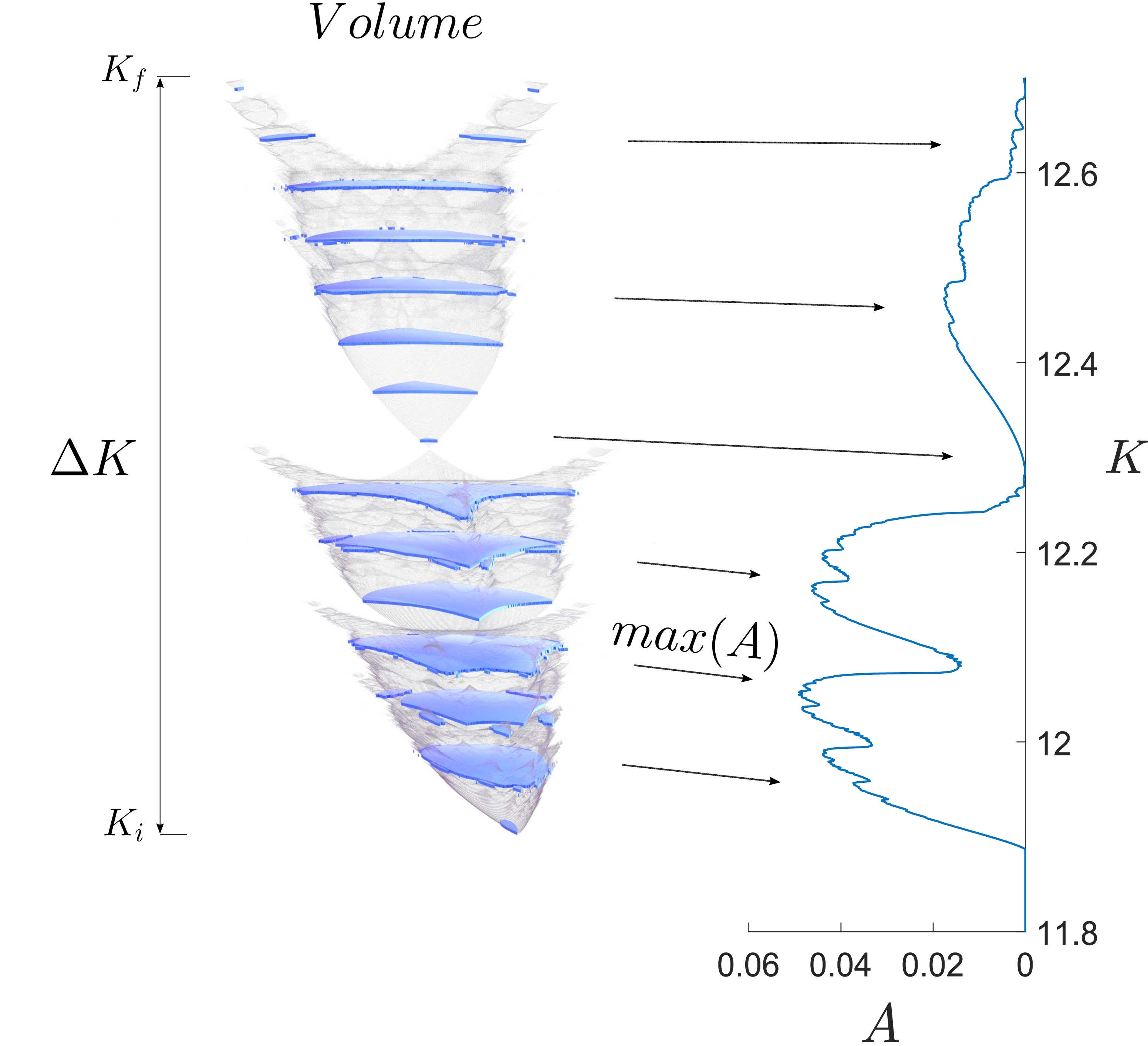}	
		\caption{Main ingredients of our method for determining the length, maximum area, and volume of the islets. The gray volume corresponds to a 3D islet, while blue areas (cross sections at constant $K$) illustrate the islet's area for various values of $K$. The blue curve on the right displays the evolution of the area in the phase space $(\theta,p)$ as a function of $K$. Some characteristic values of the area, including its maximum, are indicated with arrows. The length of the islet is $\Delta K=K_f-K_i$.}
		\label{new}
	\end{figure}

	The length, area and volume of an islet can be obtained from a 3D exit basin diagram. The main ingredients of the procedure we have followed are sketched in Fig.~\ref{new}. In this figure, the gray volume corresponds to the same type II islet plotted in Fig.~\ref{Fig5}b. On the other hand, we highlight specific cross sections for $K=$ const (indicated by the blue areas within the 3D islet), which represent islets in the phase space $(\theta,p)$. The area $A$ of an islet in the phase space evolves as $K$ increases, as shown on the right blue curve. Therefore, to characterize different islets in terms of their area, it is necessary to establish a reference point. In our case, we have chosen as a reference their maximum area, $\max(A)$. Finally, the length $\Delta K$ of an islet is simply the parameter range where it exists.
	
	According to Chirikov, the length and maximum area of the islets should follow power laws of the form
	\begin{equation}
		\Delta K \propto K^{a_1}, \kern 2pc \max (A) \propto K^{a_2},
		\label{powerlaw1}
	\end{equation}
	where $a_1=-1$ and $a_2=-2$. From the perspective of numerical simulations, $K$ is the parameter value where an islet appears (see $K_i$ in Fig.~\ref{new}). 
	
	As far as we know, no power laws for the volume have been suggested in the literature. However, given that the length decays as $1/K$ and the area as $1/K^2$, it leads us to conjecture that the volume should decay as $1/K^3$. Therefore, a similar power law
	\begin{equation}
		V \propto K^{a_3},
		\label{powerlaw2}
	\end{equation}
	with exponent $a_3=-3$ is expected.
	
	To show that these decay laws hold, we have studied all the islets appearing in the range of parameter values $K\in[9,500]$, where we found a total of $157$ islets ($79$ of type I and $78$ of type II). For each islet, we have computed a 3D exit basin diagram with a $1000\times1000\times1000$ resolution. 
	
\begin{figure}[h!]
		\centering
		\includegraphics[clip,height=5cm,trim=0cm 0cm 0cm 0cm]{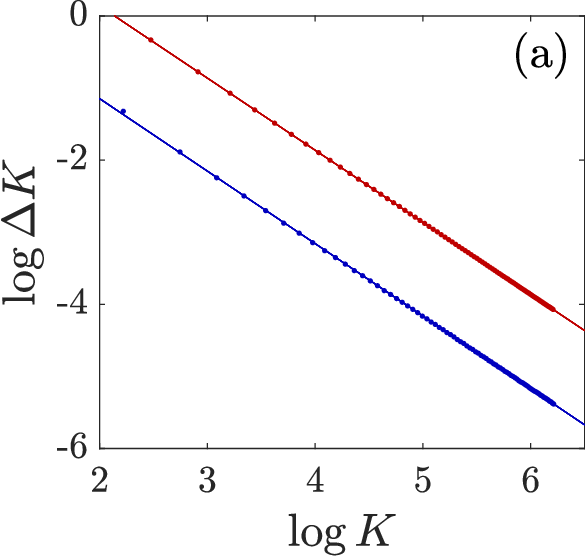}	\hspace{0.2cm}
		\includegraphics[clip,height=5cm,trim=0cm 0cm 0cm 0cm]{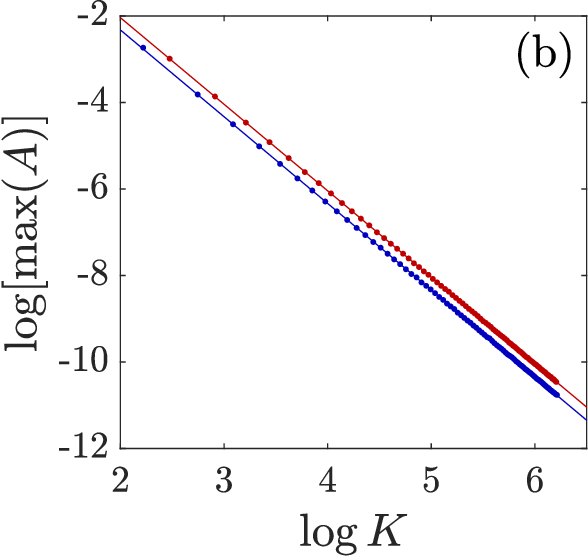} \hspace{0.2cm}	
		\includegraphics[clip,height=5cm,trim=0cm 0cm 0cm 0cm]{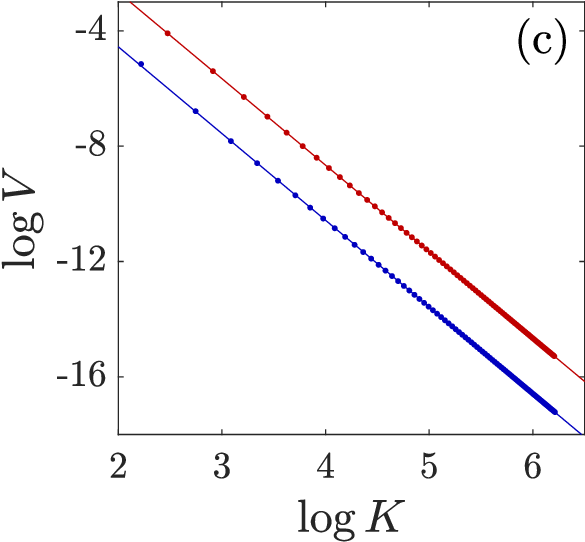}				
		\caption{Logarithmic plots for (a) length, (b) maximum area, and (c) volume of the islets. Blue dots represent type I islet data, while red dots represent type II islet data. Following the same color scheme, each data set is accompanied by a straight line obtained using the least squares method. The slope of the lines corresponds to the exponent of the power laws given by Eq.~(\ref{powerlaw1}) and Eq.~(\ref{powerlaw2}).}
		\label{Fig6}
	\end{figure}

	Once the length, maximum area, and volume have been calculated, we estimate the exponents $a_1$, $a_2$, and $a_3$ by fitting least squares lines on $\log-\log$ plots. These logarithmic plots, along with the corresponding least squares lines, are depicted in Fig.~\ref{Fig6}. Detailed results, including the decay exponents $a$, linear correlation coefficients $r$, and relative errors $\delta$, are presented in Table~\ref{T1}.
	
	\begin{table}[h!]
		\begin{tabular}{c c c c} 
			\hline		
			Fit (islet type) \kern 0.8pc & $a$ \kern 0.5pc & $r$ & $\delta$ $(\%)$   \\ [0.5ex] 
			\hline\hline
			$\Delta K$ (I) \kern 0.8pc&	 $-1.00568$ \kern 0.5pc&  $	-0.99997$ & $0.57$  \\ 
			\hline
			$\Delta K$ (II) \kern 0.8pc &	 $-1.00063$ \kern 0.5pc&  $	-0.9999991$ & $0.063$  \\ 
			\hline	
			$\max(A)$ (I) \kern 0.8pc&	 $-2.00576$ \kern 0.5pc& $-0.999992$ &  $0.29$  \\ 
			\hline
			$\max(A)$ (II) \kern 0.8pc &	$-2.00123$ \kern 0.5pc& $-0.999996$ &  $0.062$ \\ 
			\hline
			$V$ (I) \kern 0.8pc&	$-3.00883$ \kern 0.5pc& $-0.999993$ & $0.29$ \\ 
			\hline
			$V$ (II) \kern 0.8pc &	 $-2.99995$ \kern 0.5pc& $-0.99999993$ & $0.0017$ \\ 
			\hline				
		\end{tabular}
		\caption{Slope $a$ and linear correlation coefficient $r$ for the straight lines depicted in Fig.~\ref{Fig6}. The slope of the lines corresponds to the exponent of the power laws given by Eq.~(\ref{powerlaw1}) and Eq.~(\ref{powerlaw2}). The relative error $\delta$ is calculated as $|(\bar{a}-a)/\bar{a}|$, being $\bar{a}$ the theoretical value and $a$ our estimation. }
		\label{T1}
	\end{table}
	
	The experimental values match well with the theoretical predictions. Notably, the relative errors range from $0.0017\%$ to $0.57\%$. Additionally, the linear correlation coefficients are very close to $-1$ in all cases, suggesting a robust linear relationship between the variables. 
	
	\section{Conclusions and discussion}\label{sec4}
	In summary, in this paper we leverage the symmetries of the standard map to systematically identify islets of stability for a wide range of parameter values. We have accurately determined the exponents governing the power laws describing the decay of the length, maximum area, and volume of the islets. Theoretical values for the first two exponents were already established in the literature by Chirikov, but a rigorous numerical proof had not yet been provided. 
	
	In addition to these results, we have carefully described the numerical schemes used to locate islets of stability. These methods can serve as useful tools for future research endeavors aimed at identifying and studying islets in different systems.
	
	Despite their small size, islets of stability can exert a significant influence in a system. As a matter of fact, even small KAM islands can generate anomalous diffusion~\cite{Zaslasvky02,Altmann06} and modify global properties such as decay correlations \cite{Karney83}. In certain systems, maintaining stable motion may be preferable over chaotic behavior. In this situation, islets constitute a valuable subset even for large parameter values where the system is mostly (but not fully) chaotic. As a result, to find islets can be advantageous for control schemes or to define strategic parameter values. One particularly interesting potential application arises in the context of plasma confinement in tokamaks \cite{Viana23}. As an inkling of such potential application, we have conducted preliminary simulations on the Boozer map \cite{Palmero23,Punjabi}, which is a model for the configuration of the magnetic field lines of a tokamak, and we have verified that it exhibits islets of stability.

	\begin{acknowledgements} 
	This work has been financially supported by MCIN/AEI/10.13039/501100011033 and by “ERDF A way of making Europe” (Grant No. PID2019-105554GB-I00).

	\end{acknowledgements}
	
		
	\section*{Data availability}
	The data that support the findings of this study are available from the corresponding author upon reasonable request.

	\section*{Conflict of Interest}
	The authors declare that they have no conflict of interest.

\end{document}